# Whole Exome Sequencing to Estimate Alloreactivity Potential Between Donors and Recipients in Stem Cell Transplantation

Juliana K. Sampson; [1] Nihar U. Sheth; [1] Vishal N. Koparde; [1] Allison F. Scalora; [2] Myrna G. Serrano; [1] Vladimir Lee; [1] Catherine H. Roberts; [2] Maximilian Jameson-Lee; [2] Andrea Ferriera-Gonzalez; [3] Masoud H. Manjili; [4] Gregory A. Buck; [1] Michael C. Neale; [5] Amir A. Toor. [2]

*Whole Exome Sequencing in Transplant* collaborative group; Center for the Study of Biological Complexity, [1] Stem Cell Transplant Program Massey Cancer Center, [2] Department of Pathology, [3] Department of Microbiology and Immunology, [4] and the Department of Psychiatry and Statistical Genomics, [5] Virginia Commonwealth University, Richmond, VA 23298

Address correspondence to, Amir A. Toor, MD; Bone Marrow Transplant Program, Massey Cancer Center, Virginia Commonwealth University, Richmond, VA 23298.

Email: atoor@vcu.edu, Phone: 804-628-2389

Running title: Exome sequencing in stem cell transplantation

Key Statement:

1. Whole exome sequencing reveals extensive nucleotide variation outside the MHC locus in stem cell transplant donor-recipient pairs.

Abstract word count 230, manuscript word count 3132, Tables 1, Figure 5, Supplementary tables 5, Supplementary figure 1.

## Abstract

Graft versus host disease (GVHD) continues to afflict allogeneic hematopoietic stem cell transplant (SCT) recipients despite stringent HLA matching at the molecular level. In part, this is a consequence of minor histocompatibility antigen (mHA) variation between the donors and recipients resulting in alloreactivity. To determine the extent of potential antigenic variation at a molecular level, whole exome sequencing (WES) was performed on nine donor-recipient (D-R) pairs. A high frequency of sequence variation was observed between the donor and recipients' exomes independent of HLA matching. Exome variation was similar in magnitude between the recipients and their actual donors and other donors sequenced in this study, averaging 13,423 single nucleotide polymorphisms in the actual D-R pair, of which an average 6,445 were nonsynonymous. Nonsynonymous, nonconservative nucleotide variation, normalized for the number of nucleotide positions sequenced, was approximately twice as large in HLA matched unrelated compared with related D-R pairs (0.12 vs. 0.07 SNP/kbp respectively; P=0.016), indicating a greater alloreactivity potential in the former. Graft versus host and host versus graft vectors were equal in magnitude. When mapped to individual chromosomes, these polymorphic nucleotides are uniformly distributed across the entire exome. In conclusion, WES reveals extensive nucleotide sequence variation in the exomes of HLA-matched donors and recipients indicating a large potential for alloreactivity in SCT. This knowledge may guide conditioning intensity and immunosuppressive therapy administered for GVHD prophylaxis in a patient specific manner.

**Introduction**

Donor-recipient (D-R) alloreactivity may result in either graft versus host disease (GVHD) or graft rejection in allogeneic stem cell transplant (SCT) recipients, compromising outcomes. In human leukocyte antigen (HLA) matched SCT donor-recipient pairs, alloreactivity derives, in part, from minor histocompatibility antigen (mHA) differences. [1, 2] Incompatibility in mHA, caused by single nucleotide polymorphisms (SNP) in the genome, results in the recognition of recipient oligopeptides as new (non-self) antigenic epitopes by donor T cells, initiating the targeting of recipient tissues in SCT. Recipient tissue injury due to this incompatibility is characterized as GVHD.  Alternatively, loss of engraftment may occur if the recipient T cells recognize peptides of donor origin. This implies that because of the unmeasured minor histo-incompatibility between donors and recipients, outcomes in SCT remain probabilistic despite increasing stringency of HLA matching and improvements in SCT technique. [3-5]

Refinement in molecular characterization of the HLA loci has resulted in superior survival in SCT in D-R pairs matched at high-resolution. [6, 7] Thus, molecular HLA compatibility testing incorporating HLA-A, B, C, and DRB1 has become standard of care for recipients of unrelated donor allografts.  Despite this advance in histocompatibility testing, GVHD associated with D-R alloreactivity and subsequent delays in immune reconstitution remain problematic. Profiling known mHA is of limited utility in the larger context of population-based donor identification because the immunogenicity of specific mHA depends on the HLA phenotype of the patient, since individual mHA are presented efficiently only on certain HLA molecules and not on others. [8-10] Thus, donor selection algorithms in use at present leave the recipient at risk for GVHD due to the lack of knowledge of the larger antigenic landscape, as viewed from the frame of reference of the donor immune effector cells such as T and B cells.  Hypothetically, such a landscape would incorporate the information on the catalogue of mHA 'visible' to the donor and recipient T cells, and thus would represent an 'alloreactivity potential' for a D-R pair, analogous to the concept of potential energy in physics. This latent alloreactivity potential would manifest as either GVHD or graft rejection following SCT. Importantly, current conditioning regimens as well as GVHD prophylaxis regimens have been developed for use in both HLA-matched related and unrelated donor SCT in the absence of this information, which may represent an important variable in determining post transplant outcomes.

Next generation sequencing (NGS) has made it possible to comprehensively examine the role of genomic variation between donors and recipients in outcomes observed following SCT. In this report we focus on whole exome sequencing (WES), which assays only those nucleotides that code for proteins. It is likely that variation in the exome is a major source of alloreactivity because of its influence on mHA. Therefore, WES has the potential to refine donor selection algorithms by cataloging all the D-R sequence differences that may be recognized as immunogenic by either donor or recipient immune effectors.  Knowledge of the entire library of antigenic disparity may allow immunosuppressive regimens of appropriate intensity to be deployed for each D-R pair, to most effectively neutralize the alloreactivity potential in an individualized manner to optimize the SCT outcomes. In this paper the findings of a pilot study examining the extent of variation in the exomes of HLA matched SCT donor-recipient pairs is reported, identifying frequent sequence variation between the two, and therefore a large alloreactivity potential.

**Methods.**

*Patients*

Patients with recurrent hematological malignancies in complete or partial remission were enrolled on a prospective clinical trial approved by the Virginia Commonwealth University Institutional Review Board (VCU-IRB) (Clinicaltrials.gov identifier: NCT00709592). The study is a randomized phase II trial of reduced intensity conditioning regimen incorporating anti-thymocyte globulin (ATG) and 450 cGy total body irradiation for allogeneic SCT in patients with recurrent hematological malignancies. [11, 12] Post transplant immunosuppression included tacrolimus and mycophenolate mofetil. High resolution HLA matching was performed at HLA-A, B, C and DRB1 loci in 4 matched related donor (MRD) pairs and 5 matched unrelated donor (MUD) pairs, with 7/8 or 8/8 matching required for transplant eligibility. Patient characteristics and clinical outcomes are given in supplementary table 1.

*Whole exome sequencing*

Approval was obtained from the VCU-IRB to retrospectively acquire cryopreserved pre-transplant DNA samples from patients enrolled on this trial, and their donors to perform whole exome sequencing. For this analysis the DNA samples were de-identified and coded. TruSeq exome enriched libraries were prepared from the de-identified, donor-recipient pair DNA samples following standard Illumina protocol. Donor and recipient sequences were compared with each other to identify all the SNPs in the D-R pair. WES and SNP library generation protocol, and methodology for comparison between the donors and recipients is outlined in Table 1.

*Determining exome differences between donors and recipients*

The annotated SNP differences between donor and recipient samples were coded according to functionality as being either synonymous or nonsynonymous, and amongst the latter as either conservative or nonconservative or stop. To correlate the donor-recipient exome sequence difference with the risk of clinical outcomes such as GVHD incidence, the vector of the change was analyzed with respect to the donor. If the recipient sample contained a polymorphism not present in the donor, the SNP at that position was counted as being in the graft versus host (GVH) direction. Reciprocally, if the recipient sample did not contain a polymorphism present in the donor this was counted as being in the host versus graft direction (HVG) (*manuscript in preparation*). [13] The total counts for functional SNPs per pair are reported. To normalize the results based on the input data, the chromosomal positions common between two samples, regardless of presence or absence of polymorphism, were determined and then used to calculate a normalized SNP count per functional group per pair based on the following equation:

$$Normalized\ SNP\ count\ = \left( \frac{Total\ No.\ SNPs}{Total\ No.\ Common\ Positions\ /\ kbp} \right)$$

Normalization of the data allows direct comparison of different donor-recipient pairs since the number of chromosomal positions sequenced is slightly different for each sample and thus, only those chromosomal positions that were sequenced in both the donor and the recipient were considered. The Mann-Whitney U test was used to compare the degree of donor-recipient sequence difference between HLA matched related and unrelated D-R pairs.

**Results.**

*Marked whole exome sequence difference between HLA matched donors and recipients*

Whole exome enriched libraries were prepared from the nine D-R pair DNA samples (Supplementary Table 2). The D-R alloreactivity potential was estimated by quantifying the sequence variation between donors and recipients, with the initial examination of the donor-recipient exome differences focused on the major histocompatibility locus. The HLA region exome sequences of each recipient was compared with their actual donors and also with donors from the other D-R pairs sequenced, in a simulated-matching analysis. Multiple SNPs were observed across these loci in the D-R pairs, however the average number of polymorphisms was 4-10 fold higher between the recipient and simulated donors, as opposed to the actual donors (Figure 1A). This demonstrates that, as expected, matching HLA-A, B, C and DR antigens by conventional techniques reduces D-R sequence differences across the entire HLA region. Notably, the magnitude of this reduction was greater in related donors as opposed to unrelated donors, consistent with the notion that MUD will have greater variation in the non-HLA coding region of the MHC locus. Further, the one patient with HLA-A antigen mismatched related donor (patient 2), had sequence differences of a similar magnitude to MUD.

WES data from the entire exome were then considered, examining all the differences at the SNP level. The average difference between the whole exome of actual HLA matched donors and recipients was large, averaging 13,423 single nucleotide polymorphisms per actual HLA matched D-R pair, of which an average 6,445 were nonsynonymous. However, unlike the MHC locus where the nucleotide variation between simulated D-R pairs was greater as compared to actual D-R pairs, no substantial increase in SNP variation was observed at the exome level when recipients were compared with donors from other non-HLA matched pairs (Figure 1B). This observation implies that sequence variation across the remaining exome in SCT donors and recipients is frequent and independent of HLA matching, and may potentially contrbute to alloreactivity.

*Greater exome variation in HLA-matched unrelated donors*

Next WES differences between actual HLA-matched unrelated and related donors were further characterized to determine the *alloreactivity potential* which exists in each D-R pair. Nonsynonymous SNP frequency in each D-R pair, when normalized for the number of common bases, varied substantially between unrelated (median 0.18 SNP/Kbp nucleotides sequenced) and related (0.11 SNP/Kbp; P=0.016. Supplementary table 3) donor transplant recipients (Figure 2A). Within the nonsynonymous sequence variants it is more likely that nonconservative amino acid substitutions will lead to mHA oligopeptide conformational change and result in immunogenecity, and thus contribute to alloreactivity. When differentiated by nonsynonymous, nonconservative varaints, MUD SCT recipients once again had a higher measure of sequence variation when compared with MRD (0.12 vs. 0.07 SNP/kbp respectively; P=0.016) (Figure 2B). Further, non-conservative polymorphisms were consistently more frequent in all the pairs sequenced, with the ratio of conservative to non-conservative polymorphisms preserved across this small cohort. Thus, in this cohort of patients, greater exome variation was observed in patients with MUD SCT, with 3 of these 5 patients developing either delayed onset acute or chronic GVHD following withdrawal of immunosuppression (supplementary table 1), even though ATG had been used in the conditioning. Of the other two who did not develop GVHD, one patient with recurrent indolent lymphoma had stable mixed T cell chimerism in the absence of relapse, indicative of a potent graft vs. malignancy effect.

*Equal alloreactivity potential vectors in the GVH and HVG directions*

Polymorphisms present in the recipient and absent in the donor are more likely to result in GVHD because the donor T cells would lack tolerance to the mHA in the recipient tissue. Therefore the conservative and nonconservative nonsynonymous polymorphisms in the D-R pairs were examined with reference to their presence in either only the recipient (GVH direction) or the donor (HVG direction) or both. As with total allo-reactivity potential, the nonconservative variants in the GVH direction were more prevalent in the exomes of MUD SCT recipients (P=0.016; Figure 3A, Supplementary Table 3). Importantly, the alloreactivity potential *vectors* (frequency of polymoprhisms in either GVH, or HVG direction) were of a similar magnitude in both the GVH and HVG directions (Figure 3B, Supplementary tables 3 and 4).

When GVH or HVG vectors were examined in the MHC region between unique D-R pairs, there was however, a greater degree of variation observed, indicating that unlike whole exome sequence variation, the MHC region exome sequence may differ from prospective donor to donor, particularly when considering the case of multiple equivalently HLA matched unrelated donors. This greater interindividual variation of the vector at the HLA locus on chromosome 6**,** demonstrated a trend towards significance, with greater variation in MUD as compared with MRD (0.07 vs 0.17 nonsynonymous SNP/Kbp, P=0.063**).** However, the GVH and HVG vectors were again similar to each other in magnitude within individuals (Figure 4A and 4B, Supplementary table 5).

*Uniform distribution of SNPs across the exome*

To determine whether these polymorphisms were concentated in certain regions within the exome they were mapped to the individual chromosomes in each D-R pair. A composite figure depicting the findings demonstrates that the polymorphisms are distributed over the entire genome (Figure 5). This is true for both the GVH and HVG directions (data for HVG not shown). A close examination of the distribution of the exome varaints in this, admitedly small cohort, revealed that the frequency of occurrence of polymorphisms does not appear to be random, with a small number of genes demonstrating frequent variation and others with a logarithmically diminishing frequency (Figure S1). Further, the differences across the exome are equally as numerous as they are across the MHC region on chromosome 6p. Once again, consistent with the data depicted in figure 1, the variation is less so in the MHC region of MRD when compared with URD.

**Discussion.**

Since the early days of SCT, the criteria for donor selection have progressed from serologically matching both alleles of 3 major histocompatibility loci (6/6; HLA-A, B and DRB1) to the current standard of matching 4-5 loci at the allele level (8/8 or 10/10; with HLA-C and DQB1 added). [14, 15] The risk of GVHD in D-R pairs matched for these loci, though lower relative to those with one or more mismatches, is still substantial. [16, 17] Further evidence for the importance of genomic loci beyond the canonical HLA molecules is demonstrated by recipients, receiving a SCT from a MHC haplotype mismatched unrelated donor, who is otherwise matched at allele level for the HLA-A, B, C and DRB1 loci, having increased odds for developing GVHD, or diminished odds for relapse. [18, 19] These findings clearly show the influence of genetic variation on SCT outcome that is not accounted for by current D-R matching standards and may be addressed using NGS. [20, 21] In our patients, WES demonstrated significant variability in the MHC loci in these HLA matched D-R pairs, and more importantly it revealed extensive coding differences in the exomes of these individuals beyond the MHC region. The latter indicates the heavy and inevitable burden of histo-*incompatibility* that exists in SCT recipients, despite ever more stringent HLA compatibility testing.

Whereas the risk of disease relapse is in part determined by its biological features, [22, 23] the risk for GVHD in both its acute and chronic forms is affected by the level of genetic disparity between donors and recipients serving as a trigger for immune response. Evidence for this is derived from the frequent occurrence of GVHD and GVT responses in the gender mismatched (female donor-male recipient) setting. [24-28] These responses in HLA identical D-R pairs are mediated by differences in minor histocompatibility antigens (mHA), which are oligopeptides presented in the peptide-binding groove of the HLA molecule. [29, 30] The mHA are immunogenic and derived from commonly expressed proteins, which may vary due to nonsynonymous SNPs, microdeletions, insertions, inversions or copy number variations in the exons of the source gene. [31] They are dependent on HLA type of an individual for their immunogenicity, and though a number of HLA-restricted mHA have been described, [32-35] the entire spectrum of antigenic targets that affect risk for GVHD or graft versus tumor (GVT) responses remains unknown. Amongst MRD, mHA mismatches including CD31, HA-1 and HA-2 have been shown to predict higher rates of GVHD. [36] For unrelated donors, similar influence on clinical outcomes following

SCT have been noted when mHA mismatched donors were used. [37] However, despite the many (>30) mHA recognized, identifying unique mHA relevant in individual, HLA-matched D-R pairs remains challenging, with current work primarily focusing on probabilistic associations of clinical outcomes with MHC loci such as HLA DPB1. [38] This difficulty stems in part from the differential immunogenicity of variant mHA resulting from D-R SNP, compounded by the diversity of HLA in humans. Nonetheless, antigenic variation resulting from non-conservative, non-HLA SNP and consequent amino acid variation is critical in determining alloreactivity potential in transplantation. [39] This leads to the need for developing techniques that will give an accurate estimate of such variation.

A major advantage of WES in the transplant clinical context is that *any* variation between the donor and recipient will be accounted for, as observed in our data set. Although the impact of rare large-scale defects, such as gene deletions is not measurable using our analytic technique, the D-R exome variation noted nonetheless is substantial, particularly when comparing HLA-matched related and unrelated donors. The greater variation in MUD compared with MRD D-R pairs, was observed consistently despite increasing depth of analysis. It is to be noted that the oligopeptides resulting from exome variation may not all have equal immunogenicity; indeed some may not be immunogenic at all. This may be the result of factors such as, variable binding affinity of the 'non-self' oligopeptides to the HLA molecules in a unique D-R pair (thus varying presentation), or variable expression of the SNP bearing genes or even the lack of protein cleavage sites in the vicinity of the polymorphic locus, such that immunogenic oligo-peptides are not generated. However, the more numerous variant SNPs are in a given D-R pair, logically, the greater the probability of variant peptides being presented to donor T cells, with alloreactivity (GVHD or conversely, graft rejection) developing in the recipient following SCT. Thus the finding of greater exome variation in MUD D-R pairs both quantifies and provides a biological basis for greater alloreactivity potential when compared with MRD. Importantly, this provides rationale for investigating haplo-identical related donors in preference to single locus HLA-mismatched unrelated donors, particularly using modern conditioning and GVHD prophylaxis regimens. Further, the equivalence observed in the exome variation in the GVH and HVG in our data set, not only underscores the importance of appropriate conditioning and GVHD prophylaxis intensity in allogeneic SCT, but also highlights the value of pursuing tolerance induction strategies in solid organ transplantation.

The extensive library of SNPs observed in the exomes of HLA-matched donors and recipients, by quantifying the potential antigenic disparity between the two, also provides a partial explanation of the complex oligoclonal T cell repertoire observed following SCT. [12, 46] Based on these data we propose that the post-transplant T cell clonal expansion may be a function of the sequence variation in the donor recipient exome. However, rather than being a direct linear relationship, this will most likely be a complex interaction depending on the binding affinity of all the potential non-self oligopeptides in the D-R pair to the HLA class I and II molecules in that pair. Thus, HLA will serve as a filter for selecting out the immunogenic peptides, much as a colored filter allows only certain frequencies of white light to pass through. Consequently, the total number of SNPs leading to an immunogenic oligopeptide will be proportional to the total number of nonsynonymous SNPs. A caveat to this hypothesis however is that immune reconstitution, thus T cell repertoire following SCT is impacted by several other variables, such as the recipient microbiome, and pathogens encountered as well as the immunosuppressive therapy administered.

In this paper, WES performed on HLA-matched donor recipient pairs, revealed that in each D-R pair there exists an extensive 'library' of potentially immunogenic sequence differences, not accounted for by conventional histocompatibility testing techniques. Given the small cohort of patients examined in this study, the value of WES in donor selection algorithms remains to be determined, however, we posit that if determined in large cohorts of patients, measuring D-R alloreactivity potential by WES may lead to an improved understanding and estimation of GVHD and graft rejection risk. This will in turn help optimize immunosuppressive therapy following transplantation and maximize treatment benefit. We are assembling a larger cohort of patients to examine this question further.

**Acknowledgements.**

The authors gratefully acknowledge Ms. Cheryl Jacocks-Terell for her help in manuscript preparation. Funding for this study was provided by the VCU, Massey Cancer Center pilot project grant. Sequencing was performed in the Nucleic Acids Research Facilities, analysis was provided by the Bioinformatics Computational Core Laboratories, and performed in the computational environment of the Center for High Performance Computing, all at Virginia Commonwealth University. Dr. Neale and Mr. Sheth are supported by Commonwealth Health Research Board grant #236-11-13.

Authorship- JS, performed research, wrote the paper; NS, performed research, wrote the paper; VK, performed research, wrote the paper; AS, performed research, wrote the paper; CR, performed research, wrote the paper; MJL, performed research; MS, performed research; VL, performed research; AFG, performed research; MM, designed research, wrote the paper; GB, designed research, wrote the paper; MN, designed research, wrote the paper; AT, proposed and designed research, wrote the paper.

Conflict of Interest- Dr. Toor and Dr. Manjili have received research support from Sanofi-Aventis manufacturers of Thymoglobulin

**Table 1.** Whole exome sequencing of donor and recipient DNA, and sequence comparison to generate alloreactivity potential.

| | **TruSeq exome enriched libraries prepared from de-identified, D-R pair DNA samples Illumina protocol.** |
|---|---|
| 1 | DNA fragmentation, adapter ligation and amplification performed. |
| 2 | Libraries validated on BioAnalyzer, quantified using qPCR and pooled. |
| 3 | Exome enrichment. Two hybridizations performed using target specific biotinylated oligos followed by binding to magnetic streptavidin beads and three washes. PCR amplification of enriched product performed. Validation and sequencing on Illumina HiSeq 2000 with 4-8 samples per lane. |
| 4 | The ~100 bp paired end FASTQ reads generated by the sequencer run through the Next-generation Sequencing Quality Control (NGS QC) Toolkit [40] to select high quality (HQ) reads, i.e., reads where at least 70% of the bases had a quality score of ≥25. An average 20% reads excluded due to this HQ filtering. |
| 5 | HQ reads aligned to the Human Genome (hg18) using CLC Bio Assembly Cell version 3.22. >91% of the HQ reads aligned with at least 95% of the bases matching over 95% of the read length. The alignments converted to the industry-standard Binary sequence Alignment/Map (BAM) format. |
| 6 | SAMtools [41] used to remove PCR duplicates from the BAM files as these may bias subsequent SNP calling. All samples with at least 28X average coverage of the entire human exome, ensuring credible and accurate SNP calling. |
| 7 | SNP calling performed with preprocessed BAM files using the Broad Institute's Genome Analysis Toolkit [42] (GATKv1.6). The GATK SNP calling [43] involved three steps; 1. DNA insertion-deletion (INDEL) realignment; 2. Quality score recalibration; 3. SNP discovery and genotyping. The SNP caller generates a multi-sample VCF (variant-calls file). |
| 8 | The multi-sample VCF file filtered to remove chromosomal positions, which did not have at least 10X coverage and did not exceed 500X coverage. Insertion/deletion variants removed using VCFtools software (v.0.1.9.0). [44] |
| 9 | Each sample was separated from the multi-sample VCF file into individual files and positions containing missing genotype data removed. Since the original VCF file contained multiple samples, every alternate allele that occurred in any of the samples was represented. |
| 10 | To annotate the SNPs, the alternate allele and genotype data was transformed into an ANNOVAR-acceptable format primarily consisting of a single alternate allele and a genotype containing only combinations of zero and one. |
| 11 | Transformed data samples underwent independent comparison and annotation. 1. Recipient samples were compared to the actual donor and to every other donor sample to generate actual matches (recipient with its HLA-matched donor) and simulated donor-recipient matches (recipient with other, HLA-unmatched donor). 2. For annotation, files were first filtered to remove any positions where the genotype was the same as the reference and then annotated using ANNOVAR (v.2012Mar08). [45] |
| 12 | The sample-pair comparison files were then combined with the annotation files by comparing the variant alleles in sample 1 and sample 2, then annotating the variant position based on which sample contains the SNP. |

**Figure 1.** WES to quantify SNPs between SCT D-R pairs. (A) Locus 6p22.1 – 21.2 consisting of the MHC region, demonstrates high sequence variation between actual and simulated D-R pairs. (B) Whole exome, demonstrates extensive variation between donors and recipients independent of HLA matching. Actual D-R pairs (<span style="color:green">green bars</span> and <span style="color:red">red *</span>) depict exome variation between recipient and their actual HLA matched donors, while simulated pairs (<span style="color:blue">blue bars</span>) consist of the recipients from each pair compared with a donor from every other D-R pair. D-R pairs 3, 5, 7, 8 and 10 underwent MUD SCT (solid line), and 2, 4, 16 and 23 MRD SCT (dashed line).

A.

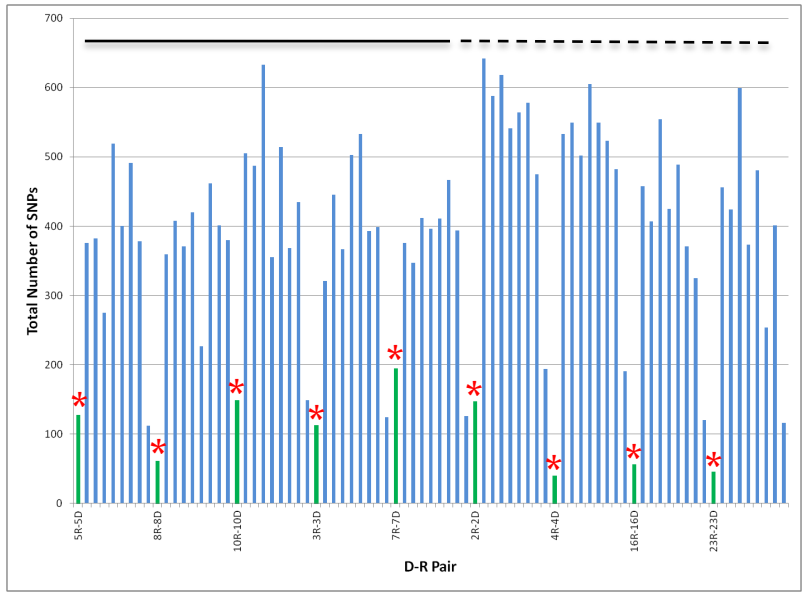

B.

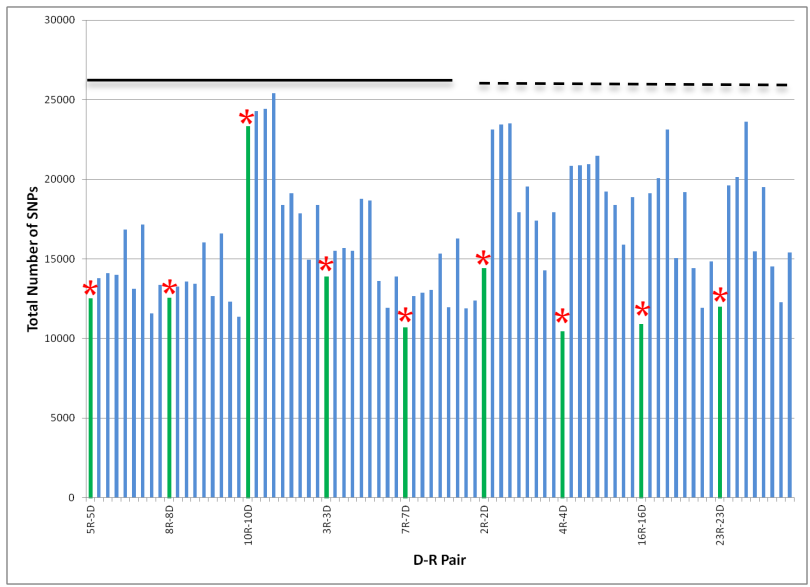

**Figure 2.** Donor-recipient alloreactivity potential. (A) Normalized data from D-R pairs depicting synonymous vs. nonsynonymous differences across whole exome.  (B) Normalized, nonsynonymous D-R exome variation accounting for conservative, nonconservative substitutions, and stop polymorphisms (either, stop-gain or stop-loss). Nonsynonymous polymorphisms are more frequent with MUD, and more likely to be non-conservative in both MRD and MUD. D-R pairs 3, 5, 7, 8 and 10 underwent MUD SCT; 2, 4, 16 and 23 MRD.

A.

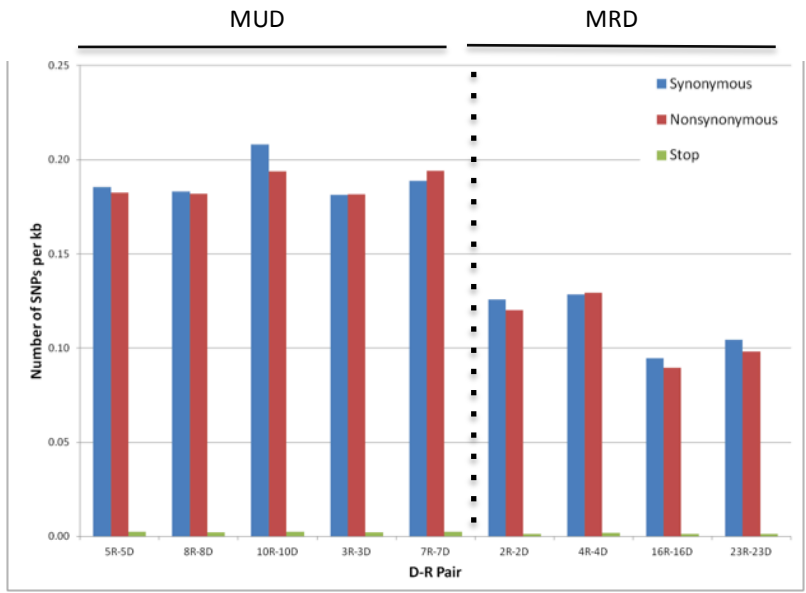

B.

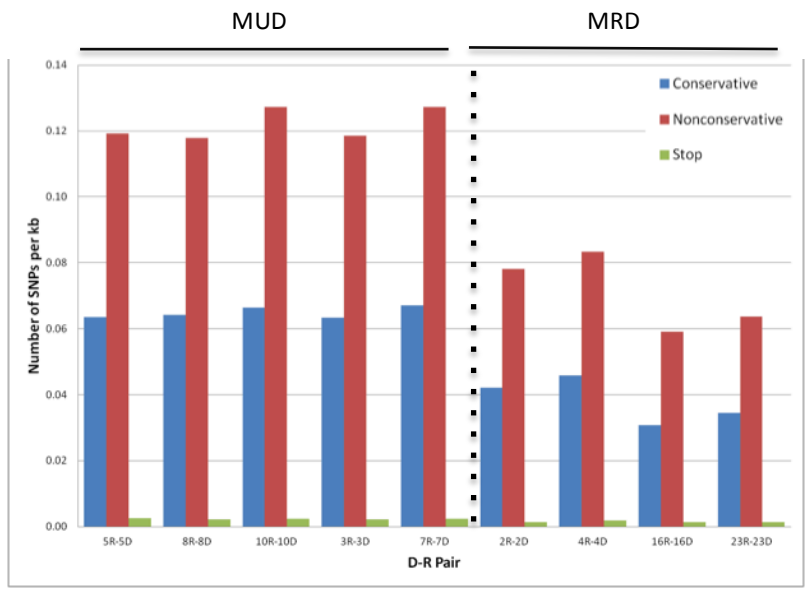

**Figure 3.** Alloreactivity potential vectors depicting conservative vs. nonconservative, nonsynonymous nucleotide variation in D-R pairs across the whole exome. (A) Graft versus host direction, variants present in the recipient and absent in the donor (B) Host versus graft direction, variants absent in the recipient and present in the donor. Changes in both directions are equal in magnitude, and of a higher magnitude in D-R pairs with an unrelated donor.

A.

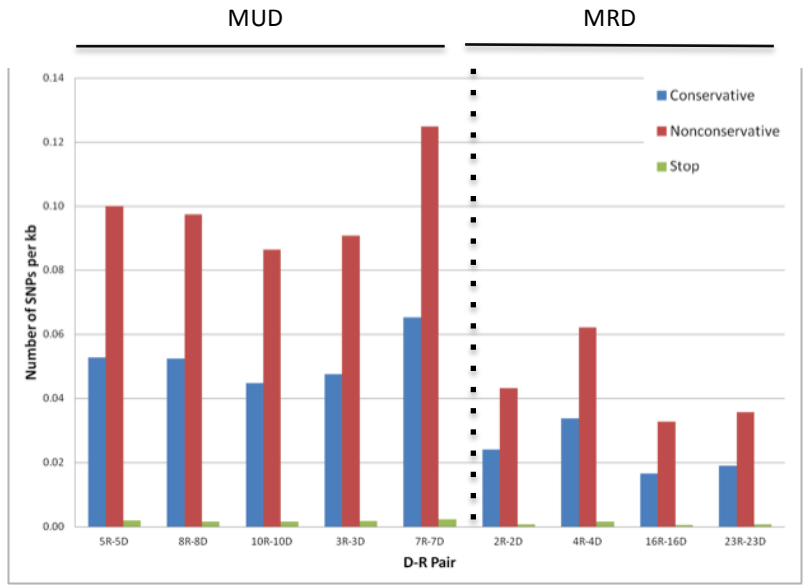

B.

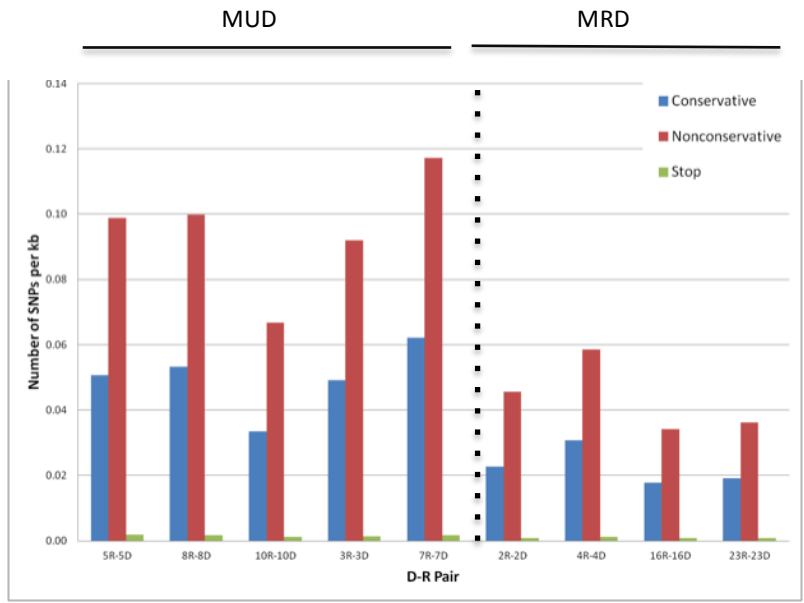

**Figure 4.** Conservative versus nonconservative, nonsynonymous nucleotide variation in D-R pairs at the 6p22.1-21.2 locus containing the MHC locus. (A) GVH direction, variants present in the recipient and absent in the donor (B) HVG direction, variants absent in the recipient and present in the donor. Greater proportional variability identified in the MHC locus exome compared with whole exome, consistent with its polymorphic nature.

A.

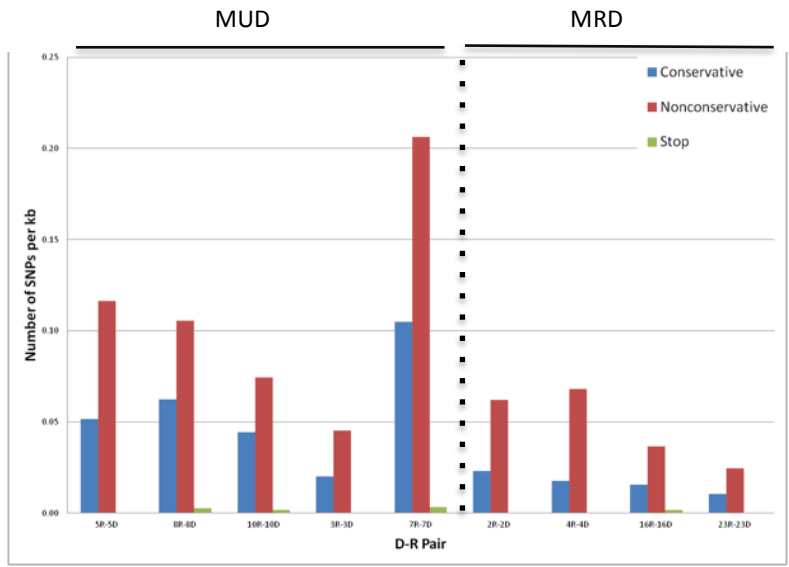

B.

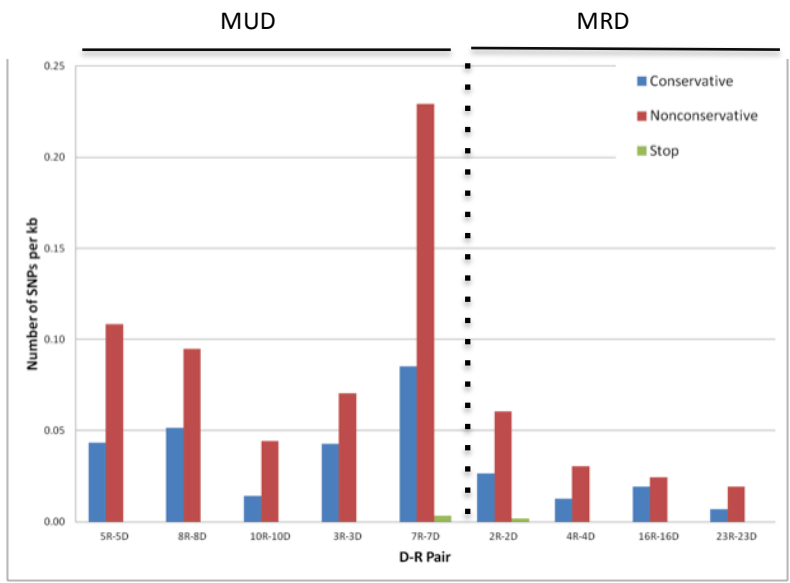

Figure 5. Nonsynonymous SNPs mapped on to individual chromosomes in all the D-R pairs, demonstrate the genomic location of polymorphisms along the length of the chromosome in each D-R pair. All nine D-R pairs depicted on the y-axis for each chromosome; SNP coordinates (location) along the length of each chromosome depicted on the x-axis. Inset shows the MHC region on chromosome 6p22. D-R pairs above dotted line (inset) from MRD with less sequence variation compared with URD. Red dots: nonsynonymous, conservative polymorphisms; blue dots: nonsynonymous, nonconservative polymorphisms; green dots: stop polymorphisms.

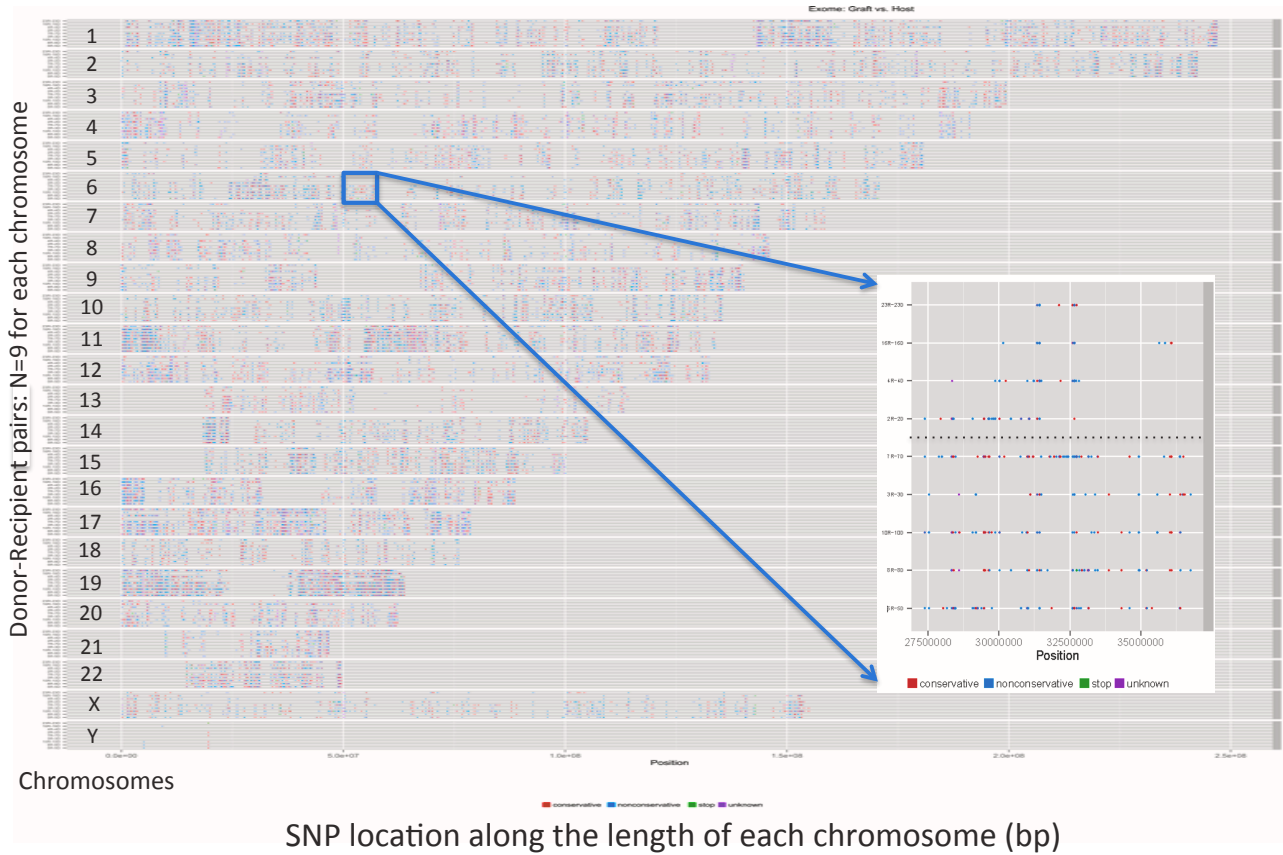

SNP location along the length of each chromosome (bp)

**References.**

# Whole Exome Sequencing to Estimate Alloreactivity Potential between Donors and Recipients in Stem Cell Transplantation

## Supplementary Materials

**Supplementary Table 1.** Donor-Recipient characteristics and patient outcomes observed in patients transplanted using ATG+450 cGy TBI.

| D-R Pair | D/R age | D/R sex | D/R race | Donor match | Diagnosis§ | GVHD | Relapse and Survival Status |
|---|---|---|---|---|---|---|---|
| 2 | 59/64 | M/M | AA/ AA | MRD* | MM | No | R, D |
| 3 | 32/44 | M/F | M/C | URD | NHL | No | A |
| 4 | 52/46 | F/F | AA/ AA | MRD | MM | No | A |
| 5 | 36/58 | M/M | C/C | URD | CLL | Yes [A] | D |
| 7 | 46/56 | M/F | C/C | URD | NHL | No | A |
| 8 | 40/57 | M/M | C/C | URD | PLL | Yes [C] | D |
| 10 | 24/57 | M/F | C/AA | URD | MM | Yes [A, C] | R, D |
| 16 | 65/62 | M/M | C/C | MRD | PLL | No | R, D |
| 23 | 58/55 | F/F | C/C | MRD | NHL | No | A |

AA indicates African American; M, Multiracial; C, Caucasian; MRD, matched related donor; URD, unrelated donor; *, 7/8 HLA mismatch (A antigen); MM- Multiple myeloma; NHL- Non-Hodgkin Lymphoma; CLL- Chronic lymphocytic leukemia; PLL- Pro-lymphocytic leukemia; §- Disease status at SCT was complete remission in NHL and CLL patients and very good partial remission in MM; [A], acute GVHD; [C], chronic GVHD; R, relapsed; D, deceased; A, alive; NA, not available.

**Supplementary Table 2.** Sample Statistics for the nine D-R pairs sequenced. The numbers in the parentheses denote percentage with respect to the number of HQ reads.

| Sample ID | No. of reads $(10^7)$ | Total Bases (MB) | No. of HQ reads $(10^7)$ | Total Bases in HQ reads (MB) | No. of reads Aligned $(10^7)$ | No. of reads with duplicates $(10^7)$ | No. of unique reads $(10^7)$ | Total Bases in unique reads (MB) | X Coverage of the Exome |
|---|---|---|---|---|---|---|---|---|---|
| 10D | 9.31 | 9307 | 7.53 | 7532 | 6.94 (92%) | 1.65 (22%) | 5.29 (78%) | 5295 | 84 |
| 10R | 11.61 | 11610 | 9.30 | 9297 | 8.47 (91%) | 2.17 (23%) | 6.29 (77%) | 6291 | 100 |
| 16D | 10.93 | 10935 | 8.92 | 8921 | 8.14 (91%) | 1.96 (22%) | 6.18 (78%) | 6181 | 98 |
| 16R | 9.67 | 9668 | 7.91 | 7911 | 7.19 (91%) | 1.62 (21%) | 5.57 (79%) | 5568 | 88 |
| 23D | 10.15 | 10153 | 8.38 | 8379 | 7.75 (93%) | 1.85 (22%) | 5.91 (78%) | 5907 | 94 |
| 23R | 10.97 | 10974 | 9.04 | 9042 | 8.32 (92%) | 2.14 (24%) | 6.18 (76%) | 6177 | 98 |
| 2D | 10.43 | 10430 | 8.37 | 8369 | 7.67 (92%) | 1.93 (23%) | 5.74 (77%) | 5738 | 91 |
| 2R | 10.95 | 10946 | 8.82 | 8822 | 8.08 (92%) | 2.09 (24%) | 5.99 (76%) | 5988 | 95 |
| 3D | 5.82 | 5819 | 4.96 | 4959 | 4.80 (97%) | 2.35 (47%) | 2.45 (53%) | 2446 | 39 |
| 3R | 6.90 | 6901 | 5.86 | 5860 | 5.67 (97%) | 2.92 (50%) | 2.75 (50%) | 2749 | 44 |
| 4D | 15.17 | 15171 | 9.53 | 9535 | 9.02 (95%) | 5.39 (57%) | 3.63 (43%) | 3631 | 58 |
| 4R | 20.13 | 20130 | 13.03 | 13034 | 12.32 (95%) | 7.93 (61%) | 4.39 (39%) | 4390 | 70 |
| 5D | 5.98 | 5976 | 5.01 | 5010 | 4.82 (96%) | 2.56 (51%) | 2.26 (49%) | 2261 | 36 |
| 5R | 5.32 | 5323 | 4.49 | 4492 | 4.33 (96%) | 2.18 (49%) | 2.15 (52%) | 2150 | 34 |
| 7D | 3.92 | 3918 | 3.32 | 3321 | 3.20 (96%) | 1.43 (43%) | 1.77 (57%) | 1766 | 28 |
| 7R | 4.39 | 4385 | 3.72 | 3717 | 3.59 (97%) | 1.68 (45%) | 1.91 (55%) | 1914 | 30 |
| 8D | 6.40 | 6401 | 5.36 | 5363 | 5.17 (97%) | 2.71 (51%) | 2.47 (50%) | 2465 | 39 |
| 8R | 5.23 | 5229 | 4.40 | 4396 | 4.24 (97%) | 2.12 (48%) | 2.12 (52%) | 2118 | 34 |

**Supplementary table 3**. Median normalized SNP/Kbp for whole exome and for the 6p22.1-21.2 (MHC) locus in the graft versus host, host versus graft and direction independent SNPs.

| | Exome | | | MHC locus | | |
|---|---|---|---|---|---|---|
| | **MRD** | **URD** | *p*-value | **MRD** | **URD** | *p*-value |
| *Graft versus Host Direction* | | | | | | |
| Synonymous | 0.065 | 0.16 | 0.016 | 0.03 | 0.11 | 0.016 |
| Nonsynonymous | 0.06 | 0.15 | 0.016 | 0.07 | 0.17 | 0.063 |
| Conservative | 0.02 | 0.05 | 0.016 | 0.02 | 0.05 | 0.032 |
| Nonconservative | 0.04 | 0.1 | 0.016 | 0.05 | 0.11 | 0.063 |
| *Host versus Graft Direction* | | | | | | |
| Synonymous | 0.065 | 0.16 | 0.016 | 0.04 | 0.1 | 0.032 |
| Nonsynonymous | 0.065 | 0.15 | 0.016 | 0.04 | 0.15 | 0.032 |
| Conservative | 0.02 | 0.05 | 0.016 | 0.015 | 0.04 | 0.111 |
| Nonconservative | 0.045 | 0.1 | 0.016 | 0.025 | 0.09 | 0.032 |
| *Direction Independent* | | | | | | |
| Synonymous | 0.115 | 0.19 | 0.016 | 0.04 | 0.13 | 0.032 |
| Nonsynonymous | 0.11 | 0.18 | 0.016 | 0.06 | 0.15 | 0.063 |
| Conservative | 0.035 | 0.06 | 0.016 | 0.025 | 0.05 | 0.063 |
| Nonconservative | 0.07 | 0.12 | 0.016 | 0.04 | 0.1 | 0.063 |

Matched related and unrelated donor SNP/Kbp distributions compared using the Mann-Whitney U Test.

MRD, matched related donor; URD, unrelated donor

**Supplementary table 4**. Total SNP counts / functional polymorphism for each D-R pair over the entire exome.

### Graft vs. Host

| Pair | Raw Counts | | | | | | Normalized Counts | | | | |
|---|---|---|---|---|---|---|---|---|---|---|---|
| | Syn | Nonsyn | Cons | Noncons | Stop | Common Positions | Syn | Nonsyn | Cons | Noncons | Stop |
| 5R-5D | 5497 | 5036 | 1737 | 3299 | 62 | 32992114 | 0.17 | 0.15 | 0.05 | 0.10 | 0.00 |
| 8R-8D | 5497 | 5013 | 1754 | 3259 | 56 | 33461071 | 0.16 | 0.15 | 0.05 | 0.10 | 0.00 |
| 10R-10D | 8218 | 7434 | 2534 | 4900 | 87 | 56634149 | 0.15 | 0.13 | 0.04 | 0.09 | 0.00 |
| 3R-3D | 5575 | 5166 | 1775 | 3391 | 66 | 37333820 | 0.15 | 0.14 | 0.05 | 0.09 | 0.00 |
| 7R-7D | 5485 | 5184 | 1779 | 3405 | 62 | 27267979 | 0.20 | 0.19 | 0.07 | 0.12 | 0.00 |
| 2R-2D | 3989 | 3838 | 1370 | 2468 | 40 | 57112850 | 0.07 | 0.07 | 0.02 | 0.04 | 0.00 |
| 4R-4D | 4099 | 3801 | 1338 | 2463 | 60 | 39689979 | 0.10 | 0.10 | 0.03 | 0.06 | 0.00 |
| 16R-16D | 3111 | 2845 | 956 | 1889 | 32 | 57757544 | 0.05 | 0.05 | 0.02 | 0.03 | 0.00 |
| 23R-23D | 3405 | 3151 | 1094 | 2057 | 38 | 57699413 | 0.06 | 0.05 | 0.02 | 0.04 | 0.00 |

### Host vs. Graft

| Pair | Raw Counts | | | | | | Normalized Counts | | | | |
|---|---|---|---|---|---|---|---|---|---|---|---|
| | Syn | Nonsyn | Cons | Noncons | Stop | Common Positions | Syn | Nonsyn | Cons | Noncons | Stop |
| 5R-5D | 5440 | 4935 | 1674 | 3261 | 61 | 32992114 | 0.16 | 0.15 | 0.05 | 0.10 | 0.00 |
| 8R-8D | 5554 | 5126 | 1784 | 3342 | 57 | 33461071 | 0.17 | 0.15 | 0.05 | 0.10 | 0.00 |
| 10R-10D | 6137 | 5676 | 1899 | 3777 | 65 | 56634149 | 0.11 | 0.10 | 0.03 | 0.07 | 0.00 |
| 3R-3D | 5608 | 5264 | 1832 | 3432 | 52 | 37333820 | 0.15 | 0.14 | 0.05 | 0.09 | 0.00 |
| 7R-7D | 5408 | 4888 | 1694 | 3194 | 48 | 27267979 | 0.20 | 0.18 | 0.06 | 0.12 | 0.00 |
| 2R-2D | 4224 | 3905 | 1301 | 2604 | 51 | 57112850 | 0.07 | 0.07 | 0.02 | 0.05 | 0.00 |
| 4R-4D | 3845 | 3540 | 1216 | 2324 | 49 | 39689979 | 0.10 | 0.09 | 0.03 | 0.06 | 0.00 |
| 16R-16D | 3065 | 3003 | 1030 | 1973 | 49 | 57757544 | 0.05 | 0.05 | 0.02 | 0.03 | 0.00 |
| 23R-23D | 3393 | 3187 | 1101 | 2086 | 55 | 57699413 | 0.06 | 0.06 | 0.02 | 0.04 | 0.00 |

### Direction Independent

| Pair | Raw Counts | | | | | | Normalized Counts | | | | |
|---|---|---|---|---|---|---|---|---|---|---|---|
| | Syn | Nonsyn | Cons | Noncons | Stop | Common Positions | Syn | Nonsyn | Cons | Noncons | Stop |
| 5R-5D | 6121 | 6024 | 2095 | 3929 | 82 | 32992114 | 0.19 | 0.18 | 0.06 | 0.12 | 0.00 |
| 8R-8D | 6125 | 6087 | 2145 | 3942 | 75 | 33461071 | 0.18 | 0.18 | 0.06 | 0.12 | 0.00 |
| 10R-10D | 11789 | 10972 | 3764 | 7208 | 136 | 56634149 | 0.21 | 0.19 | 0.07 | 0.13 | 0.00 |
| 3R-3D | 6772 | 6786 | 2365 | 4421 | 82 | 37333820 | 0.18 | 0.18 | 0.06 | 0.12 | 0.00 |
| 7R-7D | 5151 | 5297 | 1827 | 3470 | 67 | 27267979 | 0.19 | 0.19 | 0.07 | 0.13 | 0.00 |
| 2R-2D | 7191 | 6868 | 2406 | 4462 | 80 | 57112850 | 0.13 | 0.12 | 0.04 | 0.08 | 0.00 |
| 4R-4D | 5099 | 5129 | 1822 | 3307 | 74 | 39689979 | 0.13 | 0.13 | 0.05 | 0.08 | 0.00 |
| 16R-16D | 5465 | 5179 | 1771 | 3408 | 76 | 57757544 | 0.09 | 0.09 | 0.03 | 0.06 | 0.00 |
| 23R-23D | 6019 | 5668 | 1993 | 3675 | 79 | 57699413 | 0.10 | 0.10 | 0.03 | 0.06 | 0.00 |

Syn – synonymous SNPs; Nonsyn – nonsynonymous SNPs; Cons – nonsynonymous, conservative SNPs; Noncons – nonsynonymous, nonconservative SNPs; Stop – (stop gain/loss) SNPs.

**Supplementary table 5**. Total SNP counts / functional polymorphism for each D-R pair for the 6p22.1-21.2 (MHC) locus.

### Graft vs. Host

| Pair | Raw Counts | | | | | Common Positions | Normalized Counts | | | | |
|------|-----|--------|------|---------|------|---------|-----|--------|------|---------|------|
|      | Syn | Nonsyn | Cons | Noncons | Stop |         | Syn | Nonsyn | Cons | Noncons | Stop |
| 5R-5D | 41 | 62 | 19 | 43 | 0 | 369261 | 0.11 | 0.17 | 0.05 | 0.12 | 0.00 |
| 8R-8D | 64 | 62 | 23 | 39 | 1 | 369450 | 0.17 | 0.17 | 0.06 | 0.11 | 0.00 |
| 10R-10D | 38 | 67 | 25 | 42 | 1 | 564287 | 0.07 | 0.12 | 0.04 | 0.07 | 0.00 |
| 3R-3D | 28 | 26 | 8 | 18 | 0 | 397080 | 0.07 | 0.07 | 0.02 | 0.05 | 0.00 |
| 7R-7D | 72 | 95 | 32 | 63 | 1 | 305280 | 0.24 | 0.31 | 0.10 | 0.21 | 0.00 |
| 2R-2D | 22 | 48 | 13 | 35 | 0 | 563451 | 0.04 | 0.09 | 0.02 | 0.06 | 0.00 |
| 4R-4D | 17 | 34 | 7 | 27 | 0 | 396481 | 0.04 | 0.09 | 0.02 | 0.07 | 0.00 |
| 16R-16D | 9 | 30 | 9 | 21 | 1 | 573016 | 0.02 | 0.05 | 0.02 | 0.04 | 0.00 |
| 23R-23D | 13 | 20 | 6 | 14 | 0 | 573415 | 0.02 | 0.03 | 0.01 | 0.02 | 0.00 |

### Host vs. Graft

| Pair | Raw Counts | | | | | Common Positions | Normalized Counts | | | | |
|------|-----|--------|------|---------|------|---------|-----|--------|------|---------|------|
|      | Syn | Nonsyn | Cons | Noncons | Stop |         | Syn | Nonsyn | Cons | Noncons | Stop |
| 5R-5D | 37 | 56 | 16 | 40 | 0 | 369261 | 0.10 | 0.15 | 0.04 | 0.11 | 0.00 |
| 8R-8D | 34 | 54 | 19 | 35 | 0 | 369450 | 0.09 | 0.15 | 0.05 | 0.09 | 0.00 |
| 10R-10D | 30 | 33 | 8 | 25 | 0 | 564287 | 0.05 | 0.06 | 0.01 | 0.04 | 0.00 |
| 3R-3D | 50 | 45 | 17 | 28 | 0 | 397080 | 0.13 | 0.11 | 0.04 | 0.07 | 0.00 |
| 7R-7D | 55 | 96 | 26 | 70 | 1 | 305280 | 0.18 | 0.31 | 0.09 | 0.23 | 0.00 |
| 2R-2D | 46 | 49 | 15 | 34 | 1 | 563415 | 0.08 | 0.09 | 0.03 | 0.06 | 0.00 |
| 4R-4D | 18 | 17 | 5 | 12 | 0 | 396481 | 0.05 | 0.04 | 0.01 | 0.03 | 0.00 |
| 16R-16D | 16 | 25 | 11 | 14 | 0 | 573016 | 0.03 | 0.04 | 0.02 | 0.02 | 0.00 |
| 23R-23D | 11 | 15 | 4 | 11 | 0 | 573415 | 0.02 | 0.03 | 0.01 | 0.02 | 0.00 |

### Direction Independent

| Pair | Raw Counts | | | | | Common Positions | Normalized Counts | | | | |
|------|-----|--------|------|---------|------|---------|-----|--------|------|---------|------|
|      | Syn | Nonsyn | Cons | Noncons | Stop |         | Syn | Nonsyn | Cons | Noncons | Stop |
| 5R-5D | 48 | 80 | 26 | 54 | 0 | 369261 | 0.13 | 0.22 | 0.07 | 0.15 | 0.00 |
| 8R-8D | 28 | 31 | 12 | 19 | 0 | 369450 | 0.08 | 0.08 | 0.03 | 0.05 | 0.00 |
| 10R-10D | 60 | 86 | 29 | 57 | 0 | 564287 | 0.11 | 0.15 | 0.05 | 0.10 | 0.00 |
| 3R-3D | 58 | 54 | 18 | 36 | 0 | 397080 | 0.15 | 0.14 | 0.05 | 0.09 | 0.00 |
| 7R-7D | 82 | 111 | 36 | 75 | 1 | 305280 | 0.27 | 0.36 | 0.12 | 0.25 | 0.00 |
| 2R-2D | 61 | 83 | 26 | 57 | 1 | 563415 | 0.11 | 0.15 | 0.05 | 0.10 | 0.00 |
| 4R-4D | 15 | 24 | 7 | 17 | 0 | 396481 | 0.04 | 0.06 | 0.02 | 0.04 | 0.00 |
| 16R-16D | 20 | 36 | 15 | 21 | 0 | 573016 | 0.03 | 0.06 | 0.03 | 0.04 | 0.00 |
| 23R-23D | 21 | 25 | 7 | 18 | 0 | 573415 | 0.04 | 0.04 | 0.01 | 0.03 | 0.00 |

Syn – synonymous SNPs; Nonsyn – nonsynonymous SNPs; Cons – nonsynonymous, conservative SNPs; Noncons – nonsynonymous, nonconservative SNPs; Stop – (stop gain/loss) SNPs.

**Figure S1**. Distribution of variant nucleotides across specific genes, individual genes depicted on X axis with frequency of polymorphism on the y-axis (n=9 D-R pairs). (A) SNPs present in specific genes in the GVH direction. (B) SNPs present in specific genes in the HVG direction. Specific genes demonstrate varying frequency of mutation, such that, as the number of genes evaluated goes up, the number of polymorphic genes goes up logarithmically.

(A)

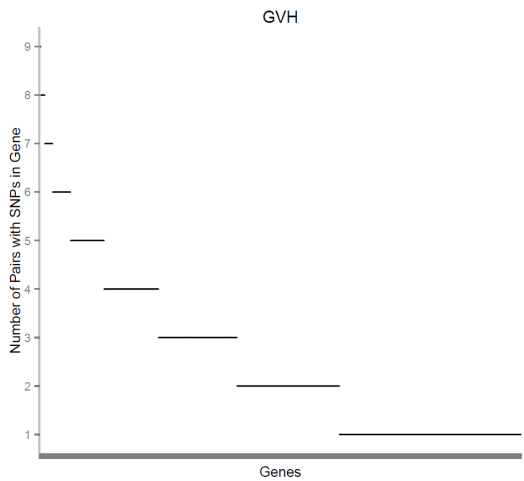

(B)

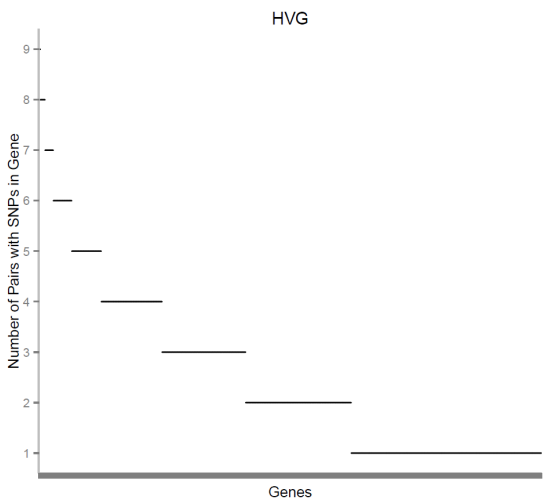